\documentclass[graybox, envcountchap]{svmult}

\usepackage{mathptmx}        
\usepackage{amsmath}
\usepackage{amssymb}
\usepackage{color}
\usepackage{helvet}          
\usepackage{courier}         
\usepackage{dirtree}

\usepackage{makeidx}        
\usepackage{graphicx}        
\usepackage{subfig}

\usepackage{multicol}        
\usepackage[bottom]{footmisc}

\usepackage{hyperref}        
\hypersetup{colorlinks=true,urlcolor=blue}

\usepackage[misc]{ifsym}

\makeindex             

\begin{document}


\title{Wind from cold accretion flows}
\author{De-Fu Bu and Feng Yuan}
\institute{De-Fu Bu (\Letter) \at Shanghai Astronomical Observatory, 80 Nandan Road, Shanghai 200030, China, \email{dfbu@shao.ac.cn}
\and Feng Yuan \at Center for Astronomy and Astrophysics and Department of Physics, Fudan University, 2005 Songhu Road, Shanghai 200438, China, \email{fyuan@fudan.edu.cn}}
%
%
\maketitle

\abstract{We describe the physical mechanisms of launching and acceleration of wind from an active galactic nucleus (AGN) accretion disk. We focus on the radiation line force driven and magnetic driven wind models, which operate on the accretion disk scale. We review the investigation histories of the two mechanisms and the most new results obtained recently. The ultra-fast outflows (UFOs) found in the hard X-ray bands are believed to directly originate from AGN accretion disks. We review the theoretical works applying the two mechanisms of wind to explain the UFOs. We briefly introduce the propagation of winds on a large scale which is important for AGN wind feedback study. Finally, the roles of wind in AGN feedback are briefly reviewed.}


\section{Introduction}
\label{sec:1}
The standard thin accretion disk (SSD, \cite{SSD:1973}) has now been widely accepted as the central engine for luminous AGNs and the high-soft state of black hole X-ray binaries. For the SSD, it is radiatively very efficient. The viscous heating rate is balanced with the radiative cooling rate. Therefore, the gas temperature is very low. The gas pressure gradient force is negligibly small compared to black hole gravity. Due to the weak vertical pressure gradient force, the disk is geometrically thin. The SSD rotates with almost Keplerian velocity and its inflow velocity is quite small. The SSD is optically thick and its spectrum is a multi-color black body.

Observations have shown that for luminous AGNs, winds with different ionization states are quite common phenomena. There is also a long history of wind study from the theoretical point of view. It is now widely accepted that the accretion process is accompanied by wind. There are three main mechanisms driving wind from an accretion disk: radiation pressure, magnetic force and thermal pressure.

For highly ionized gas, radiation pressure plays a role through Compton scattering. If an accretion system radiates with luminosity close to or higher than the Eddington luminosity, the radiation pressure can exceed black hole gravity, and winds can be launched from the accretion flow. The winds driven by radiation pressure due to Compton scattering have been intensively studied by numerical simulations (\cite{Eggum:1988} \cite{Ohsuga:2005} \cite{Yang:2014} \cite{Sadowski:2015} \cite{Sadowski:2016} \cite{Dai:2018} \cite{Hu:2022} \cite{Utsumi:2022} \cite{Yoshioka:2022}). The radiation pressure (through Compton scattering) driven wind process can occur in accretion flows both in AGNs and black hole X-ray binaries.

If the gas in an accretion disk is partially ionized, the UV photons emitted from the accretion disk can be absorbed, and the radiation pressure on resonance lines (radiation line force) can launch wind. Under proper conditions, the cross section of line resonance can be 1000 times that of Compton scattering. Therefore, radiation line force can launch wind even from sub-Eddington luminosity systems. It should be noted that radiation line force can only operate in accretion system around super-massive black holes in AGNs. For accretion disks in black hole X-ray binaries, the gas temperature is too high that radiation line force does not exist. Radiation line force driven wind from an AGN accretion disk has been studied both in analytical work \cite{Murray:1995} and numerical simulations (\cite{Proga:1998} \cite{Proga:1999} \cite{Proga:2000} \cite{Proga:2003} \cite{Yang:2021a} \cite{Yang:2021b} \cite{Nomura:2016} \cite{Nomura:2017} \cite{Nomura:2020}).

The magnetic driven wind models include magneto-centrifugal driven and magnetic pressure gradient force driven wind (see Section 1.2.2 for details). Because magnetic field must be present in black hole accretion systems which is responsible for the angular momentum transfer of the accretion disk, the magnetic driven wind models are believed to play a role in driving the wind from accretion disk in both AGNs and black hole X-ray binaries.

The wind can also be thermally driven (\cite{Begelman:1983} \cite{Woods:1996}). The X-rays generated very close to the vicinity of the black hole can heat gas on the accretion disk. If the gas can be heated up to the X-ray Compton temperature ($T_{\rm compton}$) and if $T_{\rm compton}$ is higher than the local Virial temperature, gas can be launched from the accretion disk and form wind. The value $T_{\rm compton} \sim 10^7$K, therefore, wind can roughly be launched at location $> 10^5 R_{\rm s}$ (with $R_{\rm s}$ being Schwarzschild radius), which may be beyond an accretion disk scale.

In addition to the extensive theoretical development of wind models, there are lots of observational signatures of winds from black hole accretion systems. In this Chapter, we focus on winds from luminous AGNs which are powered by cold thin accretion disks.

For AGNs, there are abundant observations of wind through blueshifted absorption lines in both UV and X-ray bands. The AGN UV absorption lines are usually classified into three groups depending on their widths, (1) the broad absorption lines (BALs), (2) the narrow absorption lines (NALs) and (3) the intermediate `mini-BALs'. The BALs are blueshifted with a velocity $v_{\rm w} \sim 10000-60000$ km s$^{-1}$, with a representative width being 10000 km s$^{-1}$. The NALs wind has velocity $v_{\rm w} \sim 100-1000 $ km s$^{-1}$. The properties of `mini-BALs' lie between those of BLAs and NALs.

Nearly $\sim 15 - 20\% $ of quasars exhibit signatures of blueshifted BALs in UV spectra (\cite{Weymann:1981} \cite{Gibson:2009}) , at transitions from ions such as Si IV $\lambda$1400, C IV $\lambda $1549, Al III $\lambda$ 1857 and Mg II $\lambda$2799 with speed $\sim $ 0.1-0.2 $c$ ($c$ is the speed of light; \cite{Hamann:2013}). The location of BALs (or the distance from the black hole) is quite difficult to be measured. In the case of BALs, one usually uses the absorption line ratios to obtain the density of BALs. Combined with the ionization parameter $\xi$ (which is defined as $\xi = L_{\rm ion}/n_{\rm e} r^2$, with $L_{\rm ion}$ being the ionizing luminosity integrated over 13.6 eV-100 keV, $n_{\rm e}$ being the electron number density and $r$ being the distance from the ionizing source) and the ionizing luminosity $L_{\rm ion}$, one can obtain the distance of BALs. Recent studies of BALs found that $50\%$ of the BALs winds are at a distance larger than $100$ pc and at least $12\%$ are at distances $> 1$ kpc (\cite{Arav:2018}). Very recently, He et al. ( \cite{He:2017}) found that the variability of the BALs wind is due to the variation of the ionizing continuum. By using the statistical properties of the variability of BALs in a sample of 915 quasars, it is found that the typical BALs wind location is tens of parsecs (\cite{He:2019}). The ionization parameter of BALs wind is $\log \xi = 0.5 - 2.5 $ erg cm s$^{-1}$, and the column density is $N_{\rm H} \sim 10^{20}-10^{23}$ cm$^{-2}$ (\cite{Laha:2021}). Theoretical models have shown that the BALs winds can have important feedback effects on their host galaxies (\cite{Hopkins:2010} \cite{Faucher:2012}). The NALs wind has a much lower velocity. The location of the NALs is $\sim 1$ pc - 1 kpc from the central black hole (\cite{Liu:2015} \cite{Laha:2021}). The ionization parameter and column density of NALs wind are $\log \xi = 0 - 1.5 $ erg cm s$^{-1}$ and $N_{\rm H} \sim 10^{18}-10^{20}$ cm$^{-2}$, respectively. Due to the low velocity, the kinetic power of NALs winds is small. Therefore, the NALs winds are thought to be insignificant in feedback.

In the soft X-ray band, warm absorbers (WAs) with velocity $v_{\rm w} \sim 100-2000$ km s$^{-1}$ are frequently detected (\cite{Canizares:1984} \cite{Halpern:1984} \cite{Madejski:1991} \cite{Reynolds:1997} \cite{George:1998} \cite{Laha:2014} \cite{Laha:2016} \cite{Blustin:2005}). WAs are mildly ionized with $\log \xi = -1 - 3 $ erg cm s$^{-1}$. The column density of WAs is in the range $N_{\rm H} \sim 10^{20}-10^{22.5}$. The observed location of WAs is 1-1000 pc (\cite{Blustin:2005}). Due to the low kinetic power, the WAs wind may not be sufficient for feedback. The WAs has been suggested to be a magneto-centrifugal wind (\cite{Fukumura:2018}) or thermally driven wind (\cite{Krolik:1995} \cite{Krolik:2001} \cite{Dorodnitsyn:2008} \cite{Mizumoto:2019}). Very recently, it is found that the WAs may be partially due to the collision of UFOs with the interstellar medium gas (\cite{Bu:2021}).

In the hard X-ray bands, the ultra-fast outflows (UFOs) with velocity $> 10000$ km s$^{-1}$ are observed very recently (\cite{Tombesi:2010a}). The UFOs are highly ionized and having highest column density compared to BALs, NALs and WAs (see the detailed introduction of UFOs in Section 1.3) . The UFOs are believed to be launched directly from the black hole accretion disk. The kinetic power of UFOs is in the range of $0.1-10\%$ of the Eddington luminosity $L_{\rm Edd}$ (\cite{Gofford:2015}). Therefore, UFOs may play an important role in AGN feedback.

The current observations of BALs, NALs and WAs winds indicate that they may be launched at a scale much larger than the size of an accretion disk. Only the UFOs are launched directly from accretion disk. In the following sections, we focus on the discussion of UFOs.

\section{Theoretical models of wind launching from cold accretion flows}

Both the radiation pressure and magnetic driven wind mechanisms can take place on accretion disk scale. However, the thermal mechanism takes place at a location $> 10^5 R_s$, which seems to be beyond the accretion disk scale of an AGN. In this Chapter, we focus on the radiation pressure and magnetic driven wind models.

\subsection{radiation line force driven wind}

A cold accretion disk is just partially ionized. In AGNs, the UV photons emitted by the disk can be absorbed by the gas on the disk, the radiation pressure on resonance lines can very effectively lift up gas from the disk and accelerate gas to form high velocity wind. The line force per gram in wind exposed to a UV flux $F_{\rm UV}$ is
\begin{equation}
g_{\rm line} = (\sigma_{\rm e} F_{\rm UV}/c) M(t)
\end{equation} where the first factor is the radiation pressure due to electron scattering and $\sigma_{\rm e}$ is the mass-scattering coefficient for free electrons. The effect of all the lines is accounted for by the force multiplier $M(t)$, which is a function of the optical depth parameter $t$,
\begin{equation}
t= \frac{\sigma_e \rho v_{\rm th}}{|dv_l/dl|}
\end{equation} where $\rho$ is gas density, $v_{\rm th}$ is the thermal velocity, and $dv_l/dl$ is the velocity gradient along the line of sight. The force multiplier is as follows (\cite{Castor:1975} \cite{Owocki:1998}),
\begin{equation}
    M(t) = k t^{-\alpha} \frac{(1+\tau_{\rm max})^{(1-\alpha)}-1}{\tau_{\rm max}^{(1-\alpha)}}
\end{equation}
where $k$ is proportional to the total number of lines, $\alpha$ is the ratio of optically thick to optically thin lines, $\tau_{\rm max} = t \eta_{\rm max}$, and $\eta_{\rm max}$ is a parameter determining the maximum value $M_{\rm max}$. $k$ is a function of both ionization parameter $\xi$ and gas temperature. $\eta_{\rm max}$ is a function of ionization parameter.

The value of $M_{\rm max}$ is a function of both $\xi$ and gas temperature $T$. It is found that $M_{\rm max}$ goes gradually from $\sim 2000$ to $\sim 5000$ as $\xi$ increases from 0 to $\sim 3$ and then decreases to $\sim 1$ at $\xi = 100$. Also, the line force becomes negligible if $T > 10^5 {\rm K}$ for any $\xi$ (\cite{Proga:2007}). Under proper conditions of $\xi$ and $T$, the line force can be $\sim 1000$ times higher than the radiation pressure due to free electrons scattering. Therefore, line force can very effectively launch and accelerate wind even in AGNs which have luminosity much lower than the Eddington value.

Murray et al. (\cite{Murray:1995}) analytically calculated the line force driven wind from an accretion disk around a $10^8 M_\odot$ (with $M_\odot$ being solar mass) black hole. The luminosity of the disk is assumed to be half of the Eddington value. The innermost radius from which wind is launched is $\sim 300 R_ {\rm s}$. It is assumed that there is entrained (hitchhiking) dense gas located on the inner edge of the wind, which blocks soft X-rays from the center but transmits UV photons. The hitchhiking gas keeps the wind from being strongly ionized. Wind gas is first lifted vertically up from the disk by the line resonance pressure due to UV photons locally. Then the UV photons from the central region exert line force on the wind almost radially. The wind can be accelerated to velocity $v_{\rm w} \sim  0.1 c$. The wind is nearly along the plane of the disk, with an opening angle of $\sim 5^\circ$. The line force driven wind is used to explain the origin of the broad absorption lines seen in some quasars. It is found that the covering fraction of the wind is $\sim 10\%$, which naturally explains the fraction of quasars showing broad absorption lines. In addition, the wind generates broad emission lines, therefore, the emitting gas in broad emission line regions may be line force driven wind.

Two dimensional axial-symmetric numerical simulations of line force driven wind were first performed in the late 1990s (\cite{Pereyra:1997} \cite{Proga:1998} \cite{Proga:1999}). We note that these simulations are designed to study the wind from cataclysmic variables (CVs). Therefore, in their simulations, the central object is a star with mass and radius appropriate to a white dwarf. Both the central star and the surrounding cold thin accretion disk can emit photons which exert line force on wind. We also mention that gas in these simulations was assumed to be globally isothermal. A dense, nearly equatorial, low-velocity wind is found which is bounded by a low density, high velocity wind in a channel at larger angles.

Numerical simulations of line force driven wind from an AGN accretion disk were carried out by Proga et al. (\cite{Proga:2000}). In these simulations, the gas temperature is self-consistently calculated by taking into account radiative cooling/heating processes. The radiative heating processes include Compton heating and photoionization heating. The radiative cooling processes include Compton cooling, recombination cooling, bremsstrahlung and line coolings. The inner radial boundary of these simulations is located at 300 $R_{\rm s}$. They found that the accretion disk atmosphere can `shield' itself from external X-rays so that the gas on the accretion disk can not be over-ionized. The local disk radiation can launch gas off the disk photosphere by radiation line force. It is also found that for a $10^8 M_\odot$ with luminosity of 0.5 $L_{\rm Edd}$, an equatorial wind can be launched from $\sim 10^{16} $cm from the black hole with a covering factor being $\sim 0.2$. The highest velocity of wind is $0.05c$ and the mass loss rate of wind is $0.5 M_\odot$ yr$^{-1}$. We note that the innermost radius at which wind is launched found by Proga et al. (\cite{Proga:2000}) roughly equals to the inner radial boundary of the simulation. Therefore, the strength of wind may be underestimated, since the wind from radius smaller than $300R_{\rm s}$ has not been simulated in the simulations. In order to capture as much as possible of the line force driven wind, Proga \& Kallman (\cite{Proga:2004}) decrease the inner radial boundary of the simulations to 30 $R_{\rm s}$, the new simulations confirm the main results from Proga et al. (\cite{Proga:2000}).

Very recently, Nomura et al. (\cite{Nomura:2016} \cite{Nomura:2017}) re-perform simulations of the line force driven accretion disk wind. They study the dependence of the properties of line force driven disk wind on black hole mass ($M_{\rm BH}$) and the Eddington ratio of radiation ($\epsilon$). They found that line force driven winds can be successfully launched for the range of $M_{\rm BH} = 10^{6-9} M_\odot$ and $\epsilon = 0.1-0.5$. They also found that no winds can be line force driven for AGNs with $\epsilon \lesssim 0.01$. They also apply the line force driven wind model to explain the ultra-fast outflows found in hard X-ray bands (see section 1.3.1.1 for details). The above simulations all assume that the mass accretion rate in the disk is fixed and does not decrease even in the presence of wind. Actually, in the presence of wind, the disk accretion rate should be decreased and the disk should be less UV-luminous, which may attenuate the strength of line force and affect the properties of wind. In 2020, Nomura et al. (\cite{Nomura:2020}) performed two-dimensional numerical simulations to study the line force driven accretion disk wind by taking into account the reduction of the disk mass accretion rate in response to the wind mass-loss. They found that even though the disk is less luminous compared to the case in which a constant accretion rate is assumed, the line force driven wind is still powerful. Roughly half the supplied accretion rate is ejected out in the wind for black holes with mass $10^8-10^{10} M_\odot$, accreting with luminosity at $L/L_{\rm Edd}=0.5-0.9$.

There are also simulations studying the effects of line force on the mass supply to an AGN (\cite{Proga:2007} \cite{Proga:2008} \cite{Liu:2013} \cite{Mosallanezhad:2019}). In these simulations, the authors focus on much larger scale, with the inner radial boundary located at $\sim 1000R_{\rm s}$. The outer radial boundary $\sim 10^6 R_{\rm s}$. The AGN corona emitting X-ray photons and the accretion disk emitting UV photons are both not resolved and treated as point sources in the simulations. Gas is injected into the computational domain from the outer radial boundary to mimic the gas supply to AGNs from interstellar medium (ISM). They found that strong wind can be driven by line force. The presence of wind significantly reduces the rate at which the central black hole is fed compared to the mass supply rate from ISM.

\subsection{magnetically driven wind}

There are two magnetically driven wind model, namely magnetocentrifugal force driven wind model and magnetic pressure driven wind model. The magnetocentrifugal force driven wind model (hereafter BP model) was established by Blandford \& Payne (\cite{Blandford:1982}). In this model, a large scale, sufficiently strong, open magnetic field is anchored in a thin, cold and Keplerian rotation accretion disk. In the BP model, the poloidal component of the magnetic field is at least comparable to the toroidal magnetic field, $|B_\phi/B_p| \lesssim 1$. The foot points of the magnetic field lines co-rotate with the accretion disk. For a steady state case, the angular velocity at any point of a magnetic field line equals that at the foot point. When a fluid element is loaded to a magnetic field line at a position higher than the foot point of the magnetic field line, the element will rotate with a super-Keplerian speed. When the field line has an angle $> 30^\circ$ relative to the rotational axis of the accretion disk, the component of the centrifugal force along the field line exceeds the component of gravity along the field line. The fluid element is accelerated out along the field line. An important feature of the BP model winds is that they require some assistance to launch the gas steadily from the surface of the accretion disk to the magnetic field lines. We note that numerical studies on BP winds usually do not resolve the vertical structure of the accretion disk but treat it as a boundary surface through which mass is loaded on to the magnetic field lines at a specified rate.

In the magnetic pressure driven wind model, it is believed that the toroidal magnetic field can be very quickly generated due to the differential rotation of accretion disk and built up with $|B_\phi/B_p| \gg 1$. The magnetic pressure of the toroidal magnetic field can launch a self-starting wind (\cite{Uchida:1985} \cite{Pudritz:1986} \cite{Stone:1994}). The wind is launched by the magnetic pressure gradient force. We note that in order to maintain a steady wind driven by the magnetic pressure, a steady supply of advected toroidal magnetic flux at the wind base is required. Otherwise, the winds are much likely to be a transient phenomenon. However, it is still not clear whether a steady supply of the toroidal magnetic flux can be maintained by the accretion disk differential rotation to compensate the escape of the magnetic flux in the wind.

\subsubsection{Analytical works}

The effects of magnetic field inclination angle with respect to the accretion disk on the BP model wind have been studied analytically (\cite{Cao:1994} \cite{Cao:1995}). It is found that the lower the inclination angle, the easier for wind to be accelerated. Cao (\cite{Cao:2014}) analytically includes the effects of radiation pressure due to electron scattering into the BP model. It is found that with the help of the radiation force, the mass loss rate in the wind is high, which leads to a slow wind.

The MHD equations describing the winds are coupled, partial differential equations, which are quite hard to be solved. In order to simplify the MHD equations, radial self-similarity assumption is also usually employed. For the self-similar solution, the physical variables (e.g, gas density, velocity, internal energy, magnetic field) are all assumed to have a power-law form of radius. Under these assumptions, the two-dimensional structure of wind can be solved analytically. The self-similarity BP wind model has been used to explain the observations of warm absorber in soft X-ray bands (\cite{Fukumura:2010}) and ultra-fast outflows in hard X-ray band (\cite{Fukumura:2015} \cite{Fukumura:2022}).

\subsubsection{Numerical simulations}

The cold thin disk has an aspect ratio $z/R \ll 1$, with $z$ and $R$ being disk scale height and radius in cylindrical coordinates, respectively. It is numerically challenging to simulate a global thin disk with the disk inside well resolved. In order to well resolve a thin disk and include the magnetorotational instability (MRI; \cite{Balbus:1991}) which is responsible for angular momentum transfer, one usually adopts the shearing box simulation. In the shearing box simulation, only a small patch of a thin disk is simulated with $\triangle R \ll R$ and $z \ll R$. The box is rotating with respect to the black hole with the keplerian speed $\Omega$ at the box center. The effective gravity is the tidal expansion of the effective potential (centrifugal + gravitational). The effective gravity in the radial and vertical direction are $3\Omega^2 x$ (with $x= R-R_0$) and $-\Omega^2 z$, respectively. It is clear that the gravity $g$ in radial direction is symmetric with respect to the box center $g_x = g_{-x}$. In this case, actually, one can not know which side (relative to the box center) the black hole resides. For the vertical component of gravity, it keeps increasing with height, which can only be used to the region with $z \ll R_0$. Actually, in a global case, in the region $z \gtrsim R_0$, the vertical component of gravity decreases with height.

Three-dimensional shearing box simulations have been performed to study the wind from a thin disk. In these simulations, the MRI process can be well captured. Therefore, the turbulent inside of the accretion disk is well included in the simulations. It is found that the wind is magneto-centrifugally driven (\cite{Bai:2013} \cite{Fromang:2013}), consistent with the BP model. However, the properties of wind provided by shearing box simulations should be quite different from those of wind launched from a real accretion disk. For example, due to the symmetry of the radial component of effective gravity about the box center, the winds below and above the midplane have opposite radial flowing direction. The time-averaged net wind mass flux in radial direction is zero and the direction of the wind constantly swap between radially inward and outward due to the constant flipping of the mean azimuthal field lines (\cite{Suzuki:2009} \cite{Bai:2013} \cite{Lesur:2014}). As introduced above, in the shearing box assumptions, the vertical component of the gravity becomes stronger and stronger as the vertical box size increases. It means that larger and larger energy for gas is required to climb out of that potential, while in reality the gravitational potential has a finite depth. This may result in the wind mass loss rate dropping dramatically when a taller box is used (\cite{Fromang:2013}). Although the MRI process can be well captured in shearing box simulations, the numerical setups of these simulations prevent a reliable estimate of the properties of wind from being made. Therefore, global simulations with the MRI process inside the accretion disk well captured are needed.

Global simulations of disk wind can be traced back to the 1990s. Stone \& Norman (\cite{Stone:1994}) performed two-dimensional axis-symmetric global simulations of disk wind driven from a thin disk threaded by uniform vertical magnetic field lines. In their simulations, due to the coarse resolution, the MRI in the disk interior is not captured. They found that wind can be launched from the disk by the magnetic pressure gradient force. The mass flux of wind at the outer radial boundary can even exceed the mass inflow rate there. Kato et al. (\cite{Kato:2002}) performed similar global simulations. They found that the driving mechanism of wind depends on the strength of the initial magnetic field. When the initial magnetic field is weak, the acceleration force is the magnetic pressure gradient. However, when the initial magnetic field is strong, the centrifugal force is responsible for wind acceleration which is consistent with the BP model. There are also 2D axsi-symmetric simulations in which only the wind region has been studied. In these simulations, the accretion disk is put outside the computational domain (\cite{Ouyed:1997} \cite{Krasnopolsky:1999} \cite{Krasnopolsky:2003} \cite{Porth:2010}). The accretion disk is put at the midplane boundary, which just supplies gas into the computational domain. Krasnopolsky et al. (\cite{Krasnopolsky:1999} \cite{Krasnopolsky:2003}) found  that the wind can be centrifugally accelerated through the Alfven and fast magnetosonic surfaces and collimated into cylinders parallel to the disk's axis, which is consistent with the BP model. Porth \& Fendt (\cite{Porth:2010}) found that the wind arises around 100 $R_{\rm s}$ and reaches only mildly relativistic speeds with Lorentz factor $\Gamma < 1.5$.

Magnetohydrodynamic turbulence induced by MRI is believed to be responsible for angular momentum transfer in accretion disks, which allows the mass be accreted to the central black hole. Wind is launched from the turbulent accretion disk. It is pointed out that the MRI turbulence in accretion disks could play a role in driving disk winds, particularly in mass loading to the surface region (\cite{Suzuki:2009} \cite{Suzuki:2010}). Therefore, ideally, simulations of wind should also capture the MRI process in accretion disk, which should leave an impression on the properties of wind. There have been several attempts. Suzuki \& Inutsuka (\cite{Suzuki:2014}) performed three-dimensional global simulations which resolved the the MRI process inside the disk. The disk is assumed to be locally isothermal. Initially, a weak vertical magnetic field is threading the disk. They found that after a quasi-steady state is achieved, winds can be produced intermittently. Wind is mainly accelerated by the magnetic pressure gradient forces. However, we note that the simulations have very limited radial and angular range with $\theta \in [\pi/2-0.5, \pi/2+0.5]$. It is not clear that whether wind has been sufficiently developed in the limited simulation box. It has been pointed that wind can be properly studied when the simulation domain is larger than the Alfven surface and fast magnetosonic surface (\cite{Blandford:1982}). Efforts have been made by Zhu \& Stone (\cite{Zhu:2018}), who performed simulations with a large radial dynamical range with the outer radial boundary being 3 orders of magnitude larger than the inner radial boundary. The whole angular region with $\theta \in [0, \pi]$ and $\phi \in [0, 2\pi]$ is simulated. By a artificial setting of disk temperature, the disk is set to have an aspect ratio $H/R=0.1$. A weak vertical magnetic field is threading the disk initially. It is found that the disk accretion mainly occurs at the coronal region. At the midplane, little inflow or even outflow is found. A BP style wind is fully developed. The wind passes though both the Alfven surface and fast magnetosonic surface. However, the mass flux of wind is only $0.4\%$ of the disk accretion rate. Both the Maxwell stress from MRI turbulence and wind torque contribute to disk angular momentum transfer. The contribution of wind torque to disk accretion is low, only $5\%$ of the disk accretion is due to the wind torque.

\begin{figure*}
	\includegraphics[width=0.9\textwidth]{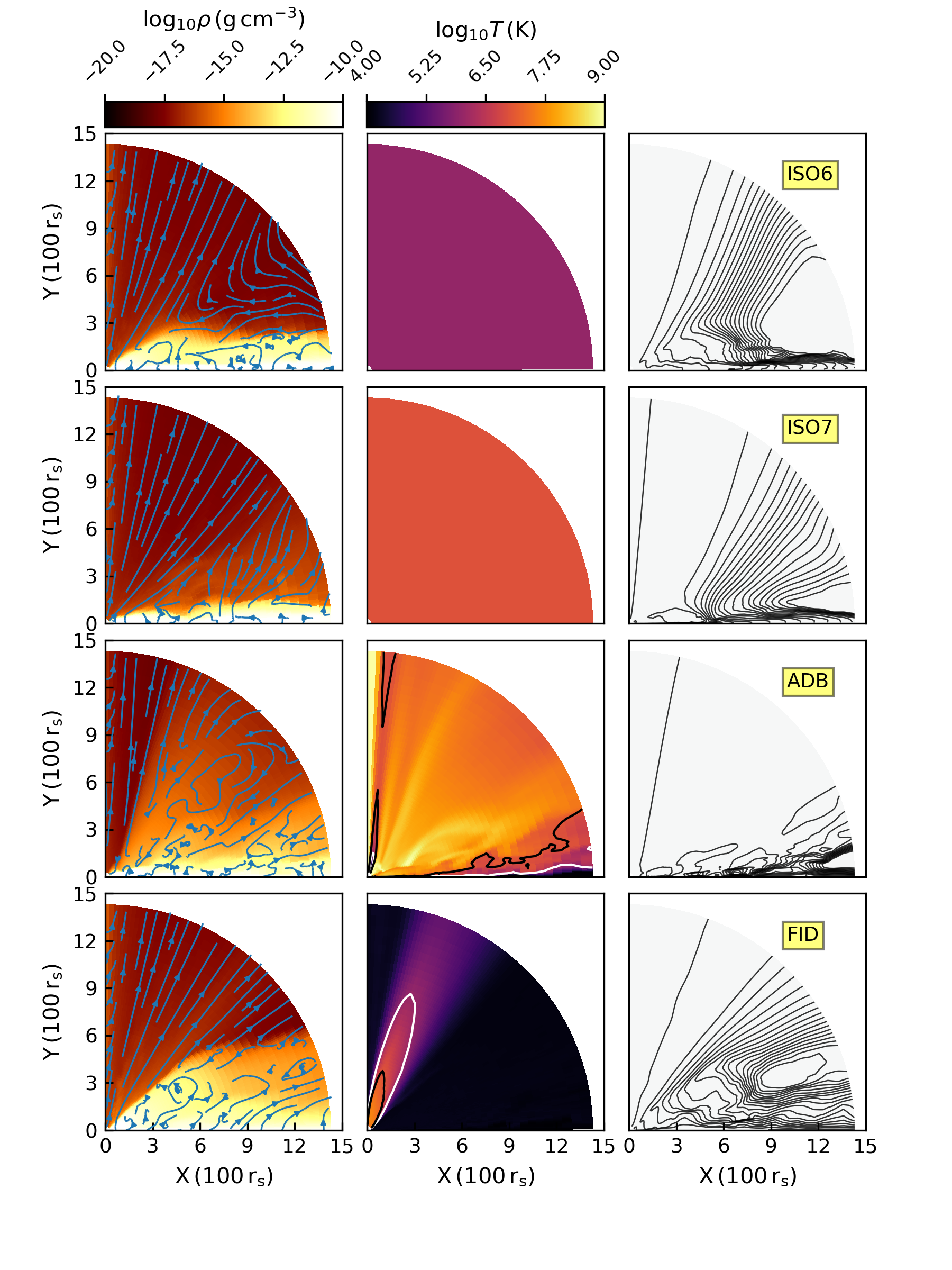}
    \caption{From left to right columns show logarithmic time-averaged density (colormap) overlaid by streamlines (blue lines with arrow), temperature (colormap) overlaid by contours with $T=10^{6,7}$K (white, black) and magnetic field lines (black contours), respectively. From top to bottom are quantities for isothermal model with $T=10^{6}$K, isothermal model with $T=10^{7}$K, adiabatic model and mole with radiative cooling/heating, respectively. It is clear that the properties of wind differ significantly from model to model. Figure from Wang et al. 2022, MNRAS, 513, 5818}
    \label{fig:timeave}
\end{figure*}

In all of the above mentioned simulations (including shearing-box and global simulations), in order to simplify the equations, the gas is usually assumed to evolve isothermally or adiabatically. However, in reality, the gas temperature should be determined by the exact radiative cooling and heating processes. It is not known whether adiabatic or isothermal assumption can be a good approximation when one studies the dynamics of wind. Wang et al. (\cite{Wang:2022}) compares the properties of wind under different assumptions of gas thermal dynamics by performing 2D simulations. In their simulations, a not resolved thin disk is put on the midplane. A Large scale open magnetic field is initially anchored on the accretion disk. They have four models, the first one assumes adiabatic evolution of gas internal energy; the second and third models assume globally isothermal gas; and the fourth one self-consistently calculates gas temperature with proper heating and cooling functions for AGNs. In the fourth model, the heating processes include the X-ray Compton scattering and photoionization heatings. The X-ray is from the AGN corona. The cooling processes include bremsstrahlung, line and recombination coolings. It is found that compared to the fourth model solving the energy equation with radiative cooling and heating, both the isothermal model and adiabatic model overestimate the temperature, underestimate the wind power, and can not predict the local structure of the winds (see Figure \ref{fig:timeave}). Therefore, it is recommended that simulations studying the properties of wind should self-consistently calculating gas temperature with proper radiative cooling and heating processes.

\subsection{Combining radiation line force and magnetic field}

The Magnetorotational instability has been shown to be a very robust and universal mechanism to produce turbulence and the transport of angular momentum in accretion disks at all radii. Therefore, magnetic fields are very likely crucial for all accretion disks. It is very likely that magnetic field plays an important role both in controlling mass accretion inside the disk and in launching winds from the disk. As introduced above, under the most favorable conditions (i.e., high UV flux, low ionization gas), the radiation line force can exceed the radiation force due to electron scattering by a factor as high as $\sim 2000$. Thus, theoretically, even in accretion disks with luminosity as low as $\sim 10^{-3} L_{\rm Edd}$, can produce strong wind. It is likely that in an AGN accretion disk both magnetic field and line force are important in driving winds. It is very important to study the two mechanisms simultaneously.

Proga (\cite{Proga:2003}) studied the hybrid model including line force driven wind and magnetic driven wind. This model is trying to study wind from nonmagnetic cataclysmic variables (nMCVs). In the model, the mass of the central object is 0.6$M_\odot$. The thin disk is not put inside the computational domain (but instead in the boundary zoos). The first active zone above the midplane represents the photosphere of the thin disk. A uniform vertical magnetic field is threading the photosphere initially. The gas is assumed to be globally isothermal with a sound speed $c_{\rm s} = 14$ km/s, which corresponds to gas temperature of $1.4 \times 10^4$ K. They found that the magnetic field deviates very quickly from the purely vertical. The toroidal component of the magnetic field grows quickly due to the differential rotation of the disk. It is also found that the toroidal field dominates over the poloidal field above the disk. Whether wind is magnetic driven or line force driven depends on the plasma beta $\beta = p_{\rm gas}/p_{\rm mag}$. When $\beta \gtrsim 1$, the wind is radiation line force driven. In this case, the wind consists of a dense, slow part that is bounded on the polar side by a high-velocity stream. The mass-loss rate is mainly due to the fast stream. As the increase of the magnetic field strength (i.e., $\beta \lesssim 1$), the dense, slow part is first affected. The slow part becomes denser and faster than its pure line force driven counterpart by a factor of $\gtrsim 100$ and $\gtrsim 3$, respectively. The dense wind begins to dominate the mass-loss rate. When the flow is strongly magnetized (i.e., $\beta \lesssim 10^{-3}$), the high-velocity stream disappears, and the wind only consists of a dense, slow outflow. The magnetic driven dense, slow wind is accelerated by the gradient of the magnetic pressure due to the toroidal field. A magnetocentrifugal wind is not found.

The hybrid model in the context of AGNs is studied recently by Yang et al. (\cite{Yang:2021a} \cite{Yang:2021b}). As done by Proga (\cite{Proga:2003}), the accretion disk itself is not simulated. The first active zoo above the midplane represents the photosphere of the thin disk. Initially, a large scale open magnetic field is threading the photosphere. There are several key differences between the work down by Proga (\cite{Proga:2003}) and Yang's works. First, in Yang's works, the central accreting object is a supermassive black hole. Second, in Yang's simulations, isothermal equation of state for gas is abandoned and the gas temperature is self-consistently calculated by taking into account of radiative heating/cooling processes. The gas is irradiated by X-rays from the AGN corona, which can heat gas. Also, gas can cool down due to radiative cooling. The magnitude of the line force sensitively depends on the ionization parameter and temperature of gas. It is very necessary to treat the gas thermal properties properly. It is found that when the magnetic field is weak, the winds are line force driven and move outward mainly in the region $\theta \gtrsim 40^\circ$, which is consistent with that found by Proga et al. in 2000 (\cite{Proga:2000}). When a relative strong magnetic field is taking into account, the vertical component of the Lorentz force can push gas into the region close to the pole, a high velocity wind can form in this region. The authors argued that the high velocity wind in the polar region can explain the UFOs in some radio-loud AGNs, which have an inclination angle of $\sim 10^\circ-70^\circ$ away from jets (\cite{Tombesi:2014}). The maximum velocity of wind found in Yang's works is around $\sim 0.1 $c. However, we note that highest velocity of the UFOs can even exceed 0.3c. The reason for a relative lower velocity found in Yang's work may be that the magnetic field is still not strong enough. Another possible reason is that the simulations are just run for 0.18 orbital times measured at the outer radial boundary, which may be too short for a higher velocity wind to develop.

\section{Comparison with observations}
\subsection{UFOs in hard X-rays band}
Recent surveys with XMM-Newton and Suzaku have shown that UFOs found through blueshifted Fe xxv and Fe xxvi K-shell absorption lines in the hard X-ray bands are observed in a significant fraction ($\sim 40-50\%$; \cite{Gofford:2015}) of active galaxies (\cite{Tombesi:2010a} \cite{Tombesi:2010b} \cite{Gofford:2013} \cite{Gofford:2015}). The outflowing velocity of the absorbers is in the range $v_{\rm w} \sim 0.03-0.3$c (\cite{Chartas:2003} \cite{Pounds:2003} \cite{Reeves:2003} \cite{Tombesi:2010a} \cite{Tombesi:2010b} \cite{Gofford:2013} \cite{Gofford:2015}). The UFOs in hard X-ray bands have been observed in both radio-quiet (\cite{Tombesi:2010a} \cite{Gofford:2013}) and radio-loud (\cite{Tombesi:2014} \cite{Gofford:2013}) AGNs. The column densities of UFOs given by photoionization modeling are on the order of ${\rm N_H} \sim 10^{22}-10^{24} {\rm cm^{-2}}$. The ionization parameter of UFOs spans $3 \lesssim {\rm \log \xi} \lesssim 6$ (\cite{Tombesi:2011}).  Given the high velocity of UFOs, it is believed that they originate from the accretion disk. On average, the UFOs are located at $r \sim 10^{15-18}$ cm (typically $\sim 10^{2-4} R_{\rm s}$) from their black hole (\cite{Gofford:2015}). The mass outflow rates are of the order $\dot M_{\rm w} \sim 0.01-1 M_\odot {\rm yr^{-1}}$ or $\sim (0.01-1) \dot M_{\rm Edd}$, kinetic power is in the range $L_{\rm w} \sim 10^{43-45} {\rm erg \ s^{-1}}$, equivalent to $(0.1-10)\%$ $L_{\rm Edd}$ (\cite{Gofford:2015}). Positive correlation between velocity of UFOs and the bolometric luminosity of its host AGNs have been found $v_{\rm w} \propto L_{\rm bol}^\beta$, with $\beta = 0.4^{+0.3}_{-0.2}$ ( \cite{Gofford:2015}). Positive relationship between the mass outflow rate of UFOs and $L_{\rm bol}$ is also found $\dot M_{\rm w} \propto L_{\rm bol}^\beta$, with $\beta = 0.9^{+0.8}_{-0.6}$, such that the UFOs in more luminous AGN are correspondingly more massive. Similar is also true for $L_{\rm w} - L_{\rm bol}$ and $\dot p_{\rm w} - L_{\rm bol}$ (with $\dot p_{\rm w}$ being the momentum flux carried by UFOs), which share strong positive slopes of $\beta = 1.5^{+1.0}_{-0.8}$ and $\beta = 1.2^{+0.8}_{-0.7}$, respectively (\cite{Gofford:2015}). Because, $L_{\rm w} \propto v_{\rm w}^3$, the observed slopes $L_{\rm w} \propto L_{\rm bol}^{\sim 1.5}$ and $\dot p_{\rm w} \propto L_{\rm bol}^{\sim 1.0}$ likely stem from their mutual dependence on $v_{\rm w}$. Gofford et al. (\cite{Gofford:2015}) argued that the velocity may be the driving factor in the observed relationships.

\subsubsection{Acceleration mechanism: radiation pressure}
The UFOs are highly ionized. Therefore, one may first recall the radiation pressure on highly ionized gas through Compton scattering.

Tombesi et al. (\cite{Tombesi:2013}) found that the momentum flux carried by UFOs roughly equals the momentum flux of the radiation field ($\dot p_{\rm rad} = L_{\rm bol}/c$) with $\dot p_{\rm w} \simeq \dot p_{\rm rad}$. The average value of the ratio is $<\dot p_{\rm w}/\dot p_{\rm rad}> = 1.6 \pm 1.1$. If UFOs are accelerated by radiation pressure due to Compton scattering, there should be a direct proportionality, $\dot p_{\rm w} \simeq C_{\rm f} \tau_{\rm e} \dot p_{\rm rad}$, where $C_{\rm f}$ is the covering factor of UFOs, $\tau_{\rm e}$ is the electron optical depth to Compton scattering. If this is the dominating acceleration mechanism, the product $C_{\rm f} \tau_{\rm e}$ should be of the order of unity. The fraction of sources with detected UFOs in the sample studied by Tombesi et al. (\cite{Tombesi:2010a}) is $\sim 0.4 - 0.6$. The average covering factor is $C_{\rm f} \simeq 0.5$. With the average column density of $N_{\rm H} \simeq 10^{23} {\rm cm^{-2}}$, one can obtain $\tau_{\rm e} \simeq \sigma_{\rm T}N_{\rm H} \simeq 0.05$ (\cite{Tombesi:2013}). The product $C_{\rm f} \tau_{\rm e}$ is much lower than unity expected from the relation $\dot p_{\rm w} \simeq \dot p_{\rm rad}$. It seems that radiation pressure due to Compton scattering is not the acceleration mechanism of UFOs. There may be one possibility that the observed UFOs are not in their acceleration phase. Therefore, there is one possibility that the observed column density is much lower than that during the acceleration phase. Through extrapolating the relation between the column density and the distance, Tombesi et al. (2013) found that at the innermost radii of $\log(r/R_{\rm s}) \sim 1$, the column density can be $N_{\rm H} = 10^{24} {\rm cm^{-2}}$. In addition, there may be presence of some ionized material that is not visible through X-ray spectroscopy. Taking all of this into account, the material may be Compton-thick with $\tau_{\rm e} \sim 1$. The product $C_{\rm f} \tau_{\rm e}$ is closer to unity.

There are several numerical simulation works, which study the origin of UFOs in the context of the line force driven wind model (\cite{Nomura:2016} \cite{Nomura:2017} \cite{Mizumoto:2021} \cite{Yang:2021a} \cite{Yang:2021b}). The UFOs are highly ionized with $3 \lesssim {\rm \log \xi} \lesssim 6$. However, as mentioned above, when $\xi > 100$, the line force can be neglected. Therefore, at first glance, it seems that the UFOs can not be line force driven. For example, Higginbottom et al. (\cite{Higginbottom:2014}) think UFOs may not be driven by radiation line force due to the high ionization and less of opacity in UV or X-ray lines. However, the situation is not such simple. At small radii, line force can lift gas up from the accretion disk photosphere. The lifted gas is illuminated by the central X-ray source. The X-ray illumination will ionize the gas, stopping the acceleration before the wind reaches the escape velocity so that the gas falls back as a failed wind. The failed wind shields outer gas from the central X-ray radiation. Therefore, at large radii, the gas can not be highly ionized. The line force can accelerate the weakly ionized gas at large radii to escape velocity before it is directly illuminated and overionized. The weakly ionized high velocity wind can not be recognized as UFOs. When the high velocity winds are exposed to X-rays and get highly ionized, they will be recognized as UFOs.

Nomura et al. (\cite{Nomura:2016}) performed hydrodynamic simulations (in spherical coordinates ($r$, $\theta$, $\phi$)) of line force driven wind and applied it to explain the origin of UFOs. In the simulations, the thin accretion disk is not resolved. The midplane ($\theta = 90^\circ$) corresponds to the surface of the accretion disk. It is assumed that the density of the disk surface ($\rho_0$) is a constant with radius. In most of the models, they set $\rho_0 = 10^{-9} {\rm g \ cm^{-3}}$. The temperature of the disk surface is set to decrease with radius $T(r) = T_{\rm in} \left(r/r_{\rm in} \right)^{-3/4}$, with $r_{\rm in} = 3R_{\rm s}$. The value $T_{\rm in}$ is set to meet the condition of $L_{\rm D} = \epsilon L_{\rm Edd} = \int^{r_{\rm out}}_{r_{\rm in}} 2\pi r \sigma T^4 dr$, with $\epsilon$ being the Eddington ratio. $r_{\rm out}$ is the outer radius of the disk where the temperature is $3\times 10^3$K. In the region $r>r_{\rm out}$, the gas temperature is too low to emit photons contributing to the line force effectively. Above the disk surface, the gas is assumed to be in hydrostatic equilibrium state. In addition to disk radiation, there is a spherical X-ray radiation source which contributes to the ionization of metals. The X-ray flux is assumed to be $10\%$ of the disk radiation. The UFOs are recognized following two conditions: (A) The outward velocity of matter with $2.5 \leq \log \xi \leq 5.5$ exceeds 10000 km s$^{-1}$. (B) The column density of matter with $2.5 \leq \log \xi \leq 5.5$ is larger than $10^{22}$ cm$^{-2}$. In their fiducial model with $M_{\rm BH} = 10^8 M_\odot$ and $\epsilon =0.5$, the UFOs are found to moving outward in the region $75^\circ \lesssim \theta \lesssim 86^\circ$. The detection probability calculated as $\Omega/4\pi$ (with $\Omega$ being the solid angle occupied by UFOs) is $20\%$. The Eddington ratio dependence of the UFO probability is calculated by performing simulations with different Eddington ratios but keeping $M_{\rm BH} = 10^8 M_\odot$. It is found that the probabilities are $17\%$, $19\%$, $20\%$, $27\%$ for $\epsilon = 0.1$, 0.3, 0.5, and 0.7, respectively. The detection probability is not sensitive to the Eddington ratio given that $\epsilon > 0.1$. It is also found that in the case $\epsilon = 0.01$, the radiation line force is too small to accelerate disk wind. Thus, the UFO probability is zero. The black hole mass dependence of the UFO probability is calculated by simulations with different black hole masses but keeping $\epsilon = 0.5$. The UFO probability just slightly decreases with increasing black hole mass: $28\%$, $22\%$, $20\%$ and $13\%$ for $M_{\rm BH}= 10^6 M_\odot$, $10^7 M_\odot$, $10^8 M_\odot$, and $10^9 M_\odot$, respectively. Tombesi et al. (\cite{Tombesi:2011}) have reported that the UFO probability to be $\sim 40\%$ based on the frequency of detecting UFOs in the Seyfert galaxies. Therefore, the UFOs detection probability given by the line force disk wind model is much smaller than that given by observations.

In 2017, Nomura and Ohsuga (\cite{Nomura:2017}) continued to study the line force driven wind. The only difference in numerical settings between this work and Nomura et al. in 2016 (\cite{Nomura:2016}) lies in the density profile of the disk surface at the midplane. In 2016, Nomural et al. (\cite{Nomura:2016}) set the density for the disk surface to be a constant with radius. However, in 2017, Nomura and Ohsuga (\cite{Nomura:2017}) set the density at the disk surface at the midplane $\theta=90^\circ$ according to the standard accretion disk model. In this paper, the authors study the dependence of the mass outflow rate, momentum flux, and kinetic power of UFOs on the disk luminosity. It is found that the observed scaling laws $\dot M_{\rm w} \propto L_{\rm bol}^{0.9}$, $L_{\rm w} \propto L_{\rm bol}^{1.5}$, and $\dot p_{\rm w} \propto L_{\rm bol}^{1.2}$ (\cite{Gofford:2015}) can roughly being explained by the line force driven wind model. However, the velocity dependence on $L_{\rm bol}$ found by the simulations is $v_{\rm w} \propto L_{\rm bol}^{1/8}$, which is different from the observed correlation of $v_{\rm w} \propto L_{\rm bol}^{1/2}$ (\cite{Gofford:2015}). It is also found that if $\epsilon \lesssim 0.01$, the line force driven wind does not appear, implying that the UFOs can not be observed in the AGNs with small Eddington ratio. Gofford et al. (\cite{Gofford:2015}) found that a large sample of UFOs is detected in AGNs with $\epsilon \sim 0.1$, and UFOs are not detected in the AGNs with $\epsilon \lesssim 0.01$. Therefore, the simulation result is consistent with observations. From Figure 5 of the work down by Nomura and Ohsuga (\cite{Nomura:2017}), we can see that the maximum velocity of line force driven wind is smaller than $0.2c$. However, observations show that the maximum velocity of UFOs can reach or even exceed $0.3c$. For example, Reeves et al. (\cite{Reeves:2018}) found an even faster Fe K UFO with speed $v_{\rm w}/c \sim 0.42$ in their 2017 \textit{XMM-Newton/NuSTAR} observation.

The simulations performed by Nomura et al. in 2016 and 2017 (\cite{Nomura:2016} \cite{Nomura:2017}) assume that the mass accretion rate is a constant with radius, although the wind mass-loss rate is a large fraction of the mass accretion rate through the disk. In 2020, Nomura et al. (\cite{Nomura:2020}) improved the calculation of the simulations so as to consider the reduction of mass accretion rate due to the wind mass loss. In the presence of wind, the mass accretion rate of the disk decreases. Therefore, the UV flux from the inner disk is reduced. However, it is found that powerful UV winds are still produced which carry enough energy and momentum to affect its host galaxy.

Based on the simulation result of Nomura et al. 2020 \cite{Nomura:2020}, Monte Carlo radiation transfer calculations have been done by Mizumoto et al. (\cite{Mizumoto:2021}) to test whether the observed UFOs can be explained by the UV line-driven disk wind model. Spectra along different lines of sight are calculated, which shows that the line driven disk wind includes sightlines where the wind is both highly ionized and fast, as required by UFOs. The PCygni-like feature in the PG1211+143 X-ray spectrum is successfully explained. This is the first time that the X-ray data from a UFO has been convincingly fit by the direct results from a UV line-driven disk wind simulation.

\subsubsection{Acceleration mechanism: Magnetocentrifugal acceleration}

Although strong correlation between the bolometric luminosity of AGNs and the kinetic power of UFOs has been detected, indicating the faster winds are more likely to be driven in higher luminosity AGNs, however, UFOs may not necessarily be radiation pressure driven (\cite{Gofford:2015}). There are some efforts trying to explain the origin of UFOs in the context of magnetocentrifugal driven wind model (\cite{Fukumura:2015} \cite{Fukumura:2018} \cite{Fukumura:2022}).

The first attempt is the application of the BP model to the UFOs found in PG 1211+143 (\cite{Fukumura:2015}). In this work, the authors first solve the steady-state axisymmetric MHD equations by assuming self-similarity. The fundamental quantity of the axisymmetric MHD equations is the magnetic stream function $\psi$, assumed to have a self-similar form $\psi (r, \theta) = (R/R_0)^q \widetilde{\psi}(\theta) \psi_o$, with $\psi_o$ as the poloidal magnetic flux through the innermost disk radius at $R_o$. $\widetilde{\psi}(\theta)$ is the angular dependence of the magnetic flux which needs to be solved, and $q$ is a free parameter which determines the radial dependence of the poloidal current. The self-similar forms of magnetic field, wind velocity, and wind number density are as follows,
\begin{equation}
\mathbf{B}(r, \theta) = (R/R_0)^{q-2} \widetilde{\mathbf{B}}(\theta) B_o
\end{equation}
\begin{equation}
\mathbf{v}(r, \theta) = (R/R_0)^{-1/2} \widetilde{\mathbf{v}}(\theta) v_o
\end{equation}
\begin{equation}
n(r, \theta) = (R/R_0)^{2q-3} \widetilde{n}(\theta) B_o^2 v_o^{-2} m_{\rm p}^{-1}
\end{equation}
where, $m_{\rm p}$ is the proton mass. The angular dependence of the variables denoted by a tilde ($\sim$) can be obtained by solving the conservation equation and the momentum equations, with the boundary values on the disk (denoted by the subscript `o') at ($R=R_o$, $\theta = 90^\circ$). With the wind mass flux, the value of $q$, the inner and outer launching radii fixed, the wind solutions can be obtained. Second, the authors calculate the ionization structure of the wind assuming the presence of a point-like ionization spectrum at the origin. The ionization spectrum includes a multicolor disk component and a X-ray power-law component. The ionizing luminosity (X-ray plus EUV) is $\sim 1.25\times 10^{44} {\rm erg \ s^{-1}}$. The photoionization code XSTAR is used, which simultaneously solves the ionization balance and thermal equilibrium of the locally illuminated wind.  It is found that the BP wind with $n \propto r ^{\sim -1}$ can explain the observed UFOs in PG 1211+143. The observed UFOs of PG 1211+143, as manifested by the Fe XXV/Fe XXVI transition properties, are determined mainly by the wind mass flux, the temperature of the ionizing spectrum and the observer's viewing angle $\theta$. It is found that under the BP model, the UFOs are mainly at a location of $\sim 1200 R_{\rm s}$, outflowing with a characteristic velocity of $v_{\rm w}/c \sim 0.1-0.2$.

The properties of UFOs vary with AGN continuum flux. For example, for PDS 456, the centroid energy of the blueshifted Fe K absorption profile increases with AGN luminosity and wind velocity v/c $\propto L_{7-30 kev}^{0.22 \pm 0.04}$, which indicates the wind is predominantly radiatively driven due to the high Eddington ratio of the source (\cite{Matzeu:2017}). Such occurences have also been found in other sources accreting at high Eddington ratio, indicating that the UFOs may be driven by intense radiation pressure in sources accreting at or close to the Eddington rate, such as in APM 08279+5255 (\cite{Chartas:2003}), PG 1211+143 (\cite{Pounds:2003}), IRAS F11119+3257 (\cite{Tombesi:2015}), 1 H 0707-495 (\cite{Hagino:2016}), and IRAS 13224-3809 (\cite{Parker:2017}).

However, we note that one can not rule out the magnetic driven origin of UFOs based on the phenomenon that the properties of UFOs vary with the the host AGN continuum flux. For example, the variations of the properties of UFOs found in the radio-quiet quasar PDS 456 were modeled in the context of the BP wind model (\cite{Fukumura:2018}). In this work, the self-similar BP wind solution is applied. Assuming plausible dependence of X-ray luminosity on the accretion rate applicable to near-Eddington state, it is found that the the photoionization calculations of the BP wind model can reproduce the observed correlations of the UFO velocity and the anti-correlation of the UFO absorber's equivalent width with the X-ray strength of PDS 456. This result indicates that even without radiation pressure, UFOs can be magnetically driven while also producing the observed spectrum and correlations with the AGN host spectrum.

\section{Propagation of wind on large-scale}
All the above mentioned theoretical works (including both analytical works and numerical simulations) have focused on the accretion disk scale. For example, the simulations studying wind usually have an outer radial boundary located at several hundreds of $R_{\rm s}$. Thus, the above works can only study the launching and propagation of winds on the accretion disk scale. Once the winds move out of the outer radial boundary of the simulations, one can not know the the evolution of wind on scales much larger than the accretion disk. However, the evolution of winds on much larger scales  is important for AGNs feedback study. Recent works have shown that winds play a significant role in the interaction between the AGNs and their host galaxies (\cite{Ciotti:2010} \cite{Ostriker:2010} \cite{Ciotti:2017} \cite{Weinberger:2018} \cite{Yuan:2018} \cite{Chen:2022} \cite{He:2022}). Therefore, we must know whether the accretion disk winds can move to galaxy or ever larger scales. After arriving at the larger scales, what will the properties of wind be?

Cui et al. (\cite{Cui:2020a} \cite{Cui:2020b}) study the dynamics of wind on large scale. In these works, it is assumed that when the wind moves at large scales, the radiation pressure exerted by the radiation from the accretion disk on wind is neglected. They also assume that the wind evolves adiabatically. They found that for wind from cold thin disk, when the Bernoulli parameter is small, it cannot propagate far, but stops at a certain radius where the Bernoulli parameter is equal to the gravitational potential energy.

\begin{figure*}
\begin{center}
\includegraphics[scale=0.3]{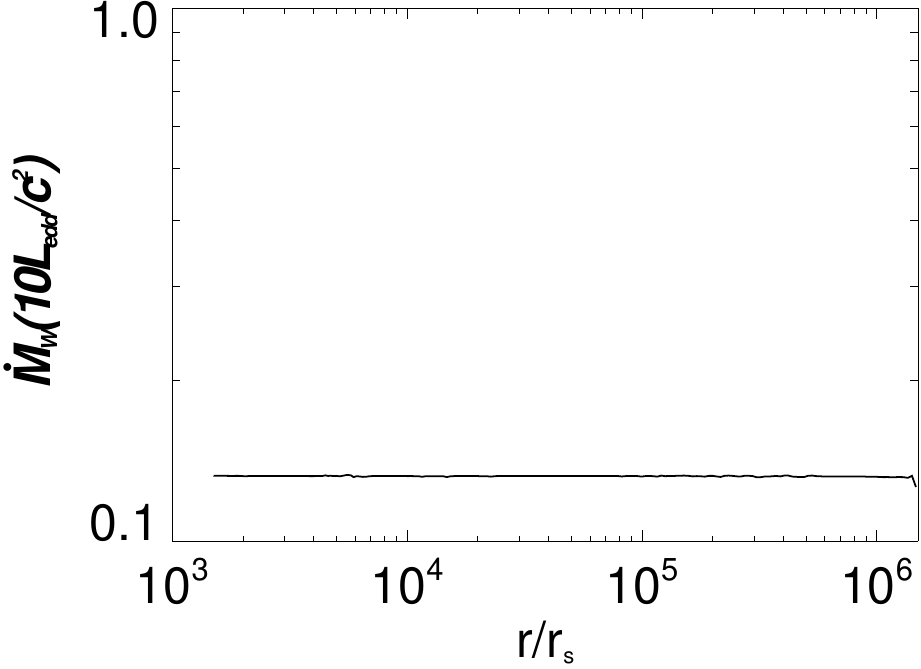}\hspace*{0.1cm}
\includegraphics[scale=0.3]{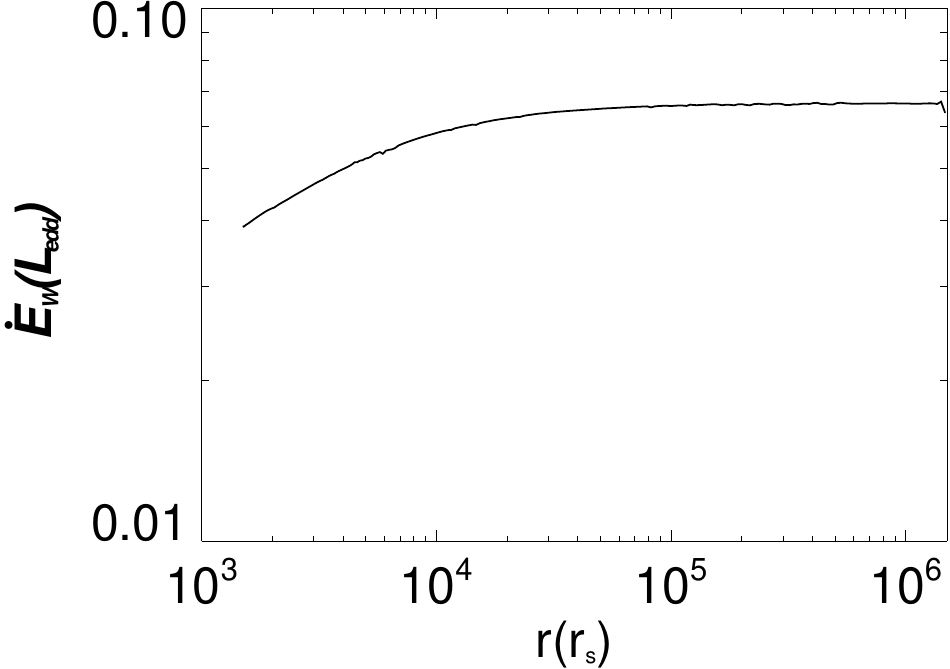}\hspace*{0.1cm}
\hspace*{0.5cm} \caption{Time-averaged radial profiles of mass flux (left panel)  and kinetic power (right panel) of winds for model with $M_{\rm BH} = 10^8 M_\odot$ and $\epsilon = 0.6$. Figrure from Zhu et al. 2022, MNRAS, 513, 1141\label{fig:2}}
\end{center}
\end{figure*}

The assumption that radiation pressure is neglected when wind propagates at large scale may be problematic. Radiation line force plays an important role in driving winds process. When winds move at large-scale, line force may continuously accelerate winds. For example, Proga (\cite{Proga:2007}) found that even at parsec scale, the radiation line force can be important. Zhu et al. (\cite{Zhu:2022}) restudied the propagation of cold line force driven wind at large-scale by numerical simulations. In these simulations, radiation line force on wind is included when wind moves at large-scale. It is found that when the disk luminosity exceeds 0.6$L_{\rm Edd}$, independent of $M_{\rm BH}$, the line force driven wind has kinetic energy flux exceeding $1\%L_{\rm Edd}$, and can escape from the black hole potential (see Figure \ref{fig:2} for an example for $M_{\rm BH} =10^8 M_\odot$ and $\epsilon = 0.6$. In this figure, the radial profiles of the mass flux and kinetic power of wind are plotted). When the disk luminosity $\sim 0.3 L_{\rm Edd}$, in the case of $M_{\rm BH} \geq 10^7 M_\odot$, winds can escape from black hole potential; however, in the case $M_{\rm BH} \leq 10^6 M_\odot$, the line force driven wind can not escape from the black hole potential.

\section{Role of cold wind in AGN feedback}
Supermassive black holes (SMBHs), with masses of $10^{6-10} M_\odot$, are universal in the center of galaxies with massive bulges. Observations show tight correlations between the black hole mass and various bulge properties, which indicate that the SMBHs co-evolves with their host galaxies ( \cite{Kormendy:2013} \cite{Heckman:2014}). The SMBH at galaxy center accretes gas from its host galaxy. In the accretion process, both radiation and outflow (including winds and jet) can be produced, which can interact with the ISM in the galaxy. The radiation can directly heat/cool the ISM. Also, radiation from the center can exert radiation pressure on ISM. Winds can directly blow away ISM. The interactions can significantly affect the properties of ISM, changing the star formation rate. In turn, the changing of properties of ISM will change the accretion circumstance of SMBHs, thus, changing the SMBH accretion rate and its growth. In this Chapter, we focus on the wind feedback.

There have been analyses about the effects of wind feedback on star formation and growth of SMBH. It has been suggested that it may be hard for the wind from the black hole accretion disk to directly couple to cod molecular gas, especially at kpc scales, the dominant reservoir for star formation (\cite{Hopkins:2010}). However, the wind can drive a wind in the hot diffuse ISM. When the hot ISM wind passes over a cold cloud, the cold gas can expand in the direction perpendicular to the incident outflow. The expansion of the cold gas may alone be enough to substantially suppress star formation in the host. Moreover, the expansion of cold gas can make itself susceptible to both ionization and momentum coupling from absorption of the incident quasar radiation field. This will drive the cold cloud at large radii to be fully ionized and driven into an outflow by radiation pressure, dramatically altering the gas properties and star formation rate in the host galaxy (\cite{Hopkins:2010}). The effects of wind feedback in an isolated elliptical galaxy have been studied (\cite{Ciotti:2010}). It is found that wind from AGN can effectively affect the inner parts ( $\sim 100$ pc) of elliptical galaxies, significantly reducing the gas infall to the central black hole. Therefore, in addition to star formation, the wind feedback can also significantly influence the black hole growth. For example, Ostriker et al. (\cite{Ostriker:2010}) show that the omission of wind feedback can lead to a hundred-fold increase in the mass of the SMBH to over $10^{10}M_\odot$. When the wind feedback is taken into account, the final SMBH mass is much lower.

Very recently, the roles of AGN feedback on the late-stage evolution of a massive elliptical galaxies are investigated by high resolution numerical simulations (\cite{Zhu:2023}). In this work, the state-of-the-art AGN physics are adopted. The AGN feedback is found to be crucial in keeping the low black hole accretion rate and suppressing the star formation. AGN wind can compensate for the radiative cooling of the gas in the galaxy. Although AGN spends most of its time in hot (radio) mode, the cumulative energy output is dominated by the outburst of the code (quasar) mode. The wind from cold mode can sweep up the gas from stellar mass-loss. The cold mode wind plays a dominant role in the late-stage evolution of massive elliptical galaxies.


\begin{acknowledgement}
The authors are supported by the Natural Science Foundation of China (grants 12173065, 12133008, 12192220, 12192223).
\end{acknowledgement}


\end{document}